\documentclass[12pt]{amsart}
\usepackage{amssymb,amsmath}
\usepackage[dvips]{graphicx}
\usepackage{pslatex}
\usepackage{amsmath,amscd}
\usepackage{amsthm}
\usepackage{overpic}
\usepackage{color}
\usepackage{hyperref}
\usepackage{tikz}
\usetikzlibrary{shapes.multipart, arrows.meta, decorations.pathreplacing}
\usepackage{tabularx}
\usepackage{float}
\usepackage{longtable}

\theoremstyle{plain} \newtheorem{thm}{Theorem}[section]

\newtheorem*{namedtheorem}{\theoremname}
\newcommand{\theoremname}{testing}

\theoremstyle{definition} \newtheorem{defn}[thm]{Definition}

\theoremstyle{remark}

\author{Christian Millichap}
\address{Department of Mathematics\\ 
Furman University\\ 
Greenville, SC 29613}
\email{christian.millichap@furman.edu}

\author{Yeeka Yau}
\address{Learning Hub (mathematics)\\ 
The University of Sydney, Australia}
\email{yeeka.yau@sydney.edu.au}

\usepackage{fancyhdr}

\fancypagestyle{plain}

\title{An artificial neural network approach to finding the key length of the Vigen\`{e}re cipher}

\begin{document}

\maketitle

\begin{abstract}
In this article, we create an artificial neural network (ANN) that combines both classical and modern techniques for determining the key length of a Vigen\`{e}re cipher. We provide experimental evidence supporting the accuracy of our model for a wide range of parameters. We also discuss the creation and features of this ANN along with a comparative analysis between our ANN, the index of coincidence, and the twist-based algorithms. 
\end{abstract}

\section{Introduction}
\label{sec:intro}

Artificial Neural Networks (ANNs) have recently seen a plethora of applications throughout academia. The development of open-source software libraries, such as TensorFlow and Keras, have made it possible to quickly train and test ANNs for a variety of purposes. In particular, there has been a steady stream of applications to cryptanalysis related questions in historical cryptology in the last $10$ years; see \cite{Ko2020} and \cite{LKELH2021} for examples of ANNs created for cipher detection among sets of classical ciphers. Here, we build an ANN to attack another important cryptanalysis question in historical cryptology: how can we accurately predict the key length of a Vigen\`{e}re cipher?

The Vigen\`{e}re cipher is one of the most well-studied historical ciphers.  The development of this cipher can be traced back to the mid 1400s and involved several cryptologists. We refer the reader to Chapter 4 of \cite{Ka1996} for a detailed description of the history of the Vigen\`{e}re cipher. For this cipher, a keyword of length $k$ is used to designate a sequence of $k$ shifts that are repeatedly used in order for encryption. Once a cryptanalyst knows the key length is  $k$, then they can partition the ciphertext into $k$  cosets, each of which contains letters that have been encrypted with the same shift (or the same alphabet for an arbitrary polyalphabetic substitution cipher). Assuming these cosets are sufficiently large, it is a straightforward task to finish breaking this cipher via frequency analysis; see \cite[Chapter 2]{Ba2021} for an example that implements this technique. This all motivates the importance of having accurate and efficient algorithms for first finding the key length. 

There are several algorithms that have been developed to predict this key length, which we review in detail in Section \ref{sec:keylength}. Historically, the Babbage--Kasiski test \cite{Ka1863} and the index of coincidence \cite{Fr1920} are the two most well known techniques, which have been applied for over a hundred years now. More recently, a series of papers have introduced the twist-based algorithms as new approaches to predict key length. The original twist algorithm was introduced by Barr--Simoson in \cite{BaSi2015}. Park--Kim--Cho--Yum provided an improved version, called the twist$^{+}$ algorithm, in \cite{PaKiChYu2020}. Further modifications were made by Millichap--Yau--Pate--Carns in \cite{MYPC2023} to build the twist$^{++}$ algorithm.  For a wide variety of key lengths and text lengths, each of these algorithms has its own strengths and weaknesses.  Therefore, it is natural to build an ANN for key length prediction that uses these algorithms in hopes of highlighting their individual strengths, while mitigating their weaknesses. 

Our work here provides an accurate model for predicting the key length of a Vigen\`{e}re cipher for a large range of key lengths and text lengths. This ANN implements both classical tools (index of coincidence in a variety of formats) and much more recent algorithms (twist-based algorithms) along with a few other features. Our exact model features are highlighted in Section \ref{subsec:CreatingModel}. Accuracy comparisons with the index of coincidence and the twist-based algorithms are given in Table \ref{fig:total_comp_graph}, which clearly highlight the superior accuracy of our ANN for a variety of text lengths. 

Our paper is organized as follows. In Section \ref{sec:keylength}, we review a variety of well-established methods from the literature for predicting the key length of a Vigen\`{e}re cipher and provide an analysis of their strengths and weaknesses. All of these tools were tested as potential features for our ANN. In Section \ref{sec:NN}, we discuss the creation, training, and evaluation of an ANN for finding the key length of the Vigen\`{e}re cipher.


\section{Key Length Attacks}
\label{sec:keylength}

In this section, we review a variety of algorithms that predict the key length of a Vigen\`{e}re cipher. Strengths and weaknesses of these algorithms will also be discussed. 


\subsection{The Babbage--Kasiski Test}
\label{BKtest}

For the Babbage--Kasiski test, one looks for repetitions of $n$-grams ($n \geq 3$) in the ciphertext. Usually, such repetitions represent the same plaintext with the same portion of the keyword used to encrypt that plaintext; see \cite{Po2006} for a analysis of accidental repetitions. Thus, the distance between these repetitions should be a multiple of the key length. By finding several such repetitions and calculating the greatest common divisor of these distances that show up most frequently, one can then formulate a conjecture for the key length. This test has the advantage of being independent of the underlying language and alphabet. 

 There are several implementation challenges that can occur when applying the Babbage--Kasiski Test. If a ciphertext does not contain any repeated trigrams or only contains a few, then the Babbage--Kasiski test might not prove helpful. In addition, the Babbage--Kasiski test could easily direct one towards a multiple of the key length or a divisor of the key length, rather than the actual key length. Furthermore, there might be multiple values that frequently show up as distinct greatest common divisor of distances between repetitions. How should one decide which is the best conjecture for the key length or an ordering for key length conjectures? 


\subsection{The Index of Coincidence}
\label{subsec:IoC}

The Index of Coincidence (IC) calculates the probability that two randomly chosen distinct letters from a text are the same. Mathematically, 

\begin{equation} IC(\mathcal{M}) = \frac{\sum_{i=1}^{26} f_i(f_i -1)}{N(N-1)}, \end{equation}

\begin{flushleft}
where $N$ is the length of a text $\mathcal{M}$ and $f_i$ represents the frequency of the $i^{th}$ letter of the alphabet in this ciphertext. When the index of coincidence is applied to a Vigen\'{e}re ciphertext $\mathcal{M}$, then one can estimate the key length $k$ via 
\end{flushleft}

\begin{equation}\label{equationkeylength} k \approx \frac{0.028N}{IC(\mathcal{M})(N-1) - 0.038N +0.066}.\end{equation}

\begin{flushleft}
Furthermore, if one has a conjectured key length of $m$, then one can partition $\mathcal{M}$ into $m$ cosets, where coset $\mathcal{M}_i$ contains the ciphertext letters encrypted by the $i^{th}$ key letter, for $1 \leq i \leq m$. From here, one can apply the IC to each $\mathcal{M}_i$. If these values approximate the $IC$ for the underlying language of the plaintext, then there is a good chance $m$ is the actual key length since the IC of a set of letters coming from a shift is the same as the IC of the corresponding plaintext letters. We refer the reader to Section 2.3 of \cite{Ba2021} for details on (2) and an example of applying the IC to cosets of a ciphertext.
\end{flushleft}

While it is simple to calculate the IC and the approximation for $k$ via Equation (2), this tool does have its weaknesses. For sufficiently long keys, estimated key lengths are sensitive to a small perturbation in the IC. Thus, the IC becomes less reliable as key length increases, which can also be seen experimentally in Figure 2 of \cite{Ma1988} and Figure 3 of \cite{PaKiChYu2020}. In addition, the key length estimate given above is dependent on your keyword containing $k$ distinct letters. If many letters are repeated in the keyword, then the IC will most likely underestimate $k$. Finally, the IC is dependent on your underlying language since different languages have different letter frequency distributions.


\subsection{The Twist-Based Algorithms}
\label{subsec:twistalg}


Before reviewing the twist-based algorithms, we first introduce some necessary notation and definitions. Suppose we are given a text $\mathcal{M}$ of length $N$. We first form a \textbf{sample signature} for $\mathcal{M}$, which is $C = < c_1, c_2, \ldots, c_{26}>$, where $c_i = \frac{f_i}{N}$ with $f_i$ representing the number of frequencies of the $i^{th}$ most common letter in $\mathcal{M}$. In other words, $C$ is the ordered set of relative frequencies for text $\mathcal{M}$. Then we can compute the \textbf{twist} of a sample signature: $$  \diamondsuit C = \displaystyle\sum_{i=14}^{26} c_i - \displaystyle\sum_{i=1}^{13}c_i.$$
When a sufficiently long text $\mathcal{M}$ is a plaintext or a ciphertext encrypted with a monoalphabetic substitution cipher, then $\diamondsuit C$ should reflect the behavior of the underlying language (English in all cases considered in this paper) and be relatively large based on the variance of frequency distributions in the underlying language. However, if $\mathcal{M}$ is a random text that would lack this variation in frequency distribution, then we should expect $\diamondsuit C$ to be quite small.  

Now, suppose $\mathcal{M}$ is a ciphertext of length $N$ that was encrypted using the Vigen\`{e}re cipher. Further, suppose we conjecture a key length of $m \in \mathbb{N}$. Then we can partition $\mathcal{M}$ into $m$ cosets $\{\mathcal{M}_j\}_{j=1}^{m}$, where $\mathcal{M}_j$ contains all the letters encrypted with the $j^{th}$ letter of the (conjectured) key of length $N$. Let $C_j$ represent the sample signature for $\mathcal{M}_j$, and let $$  \diamondsuit C_j = \displaystyle\sum_{i=14}^{26} c_{i,j} - \displaystyle\sum_{i=1}^{13}c_{i,j}$$
be the corresponding twist of each such sample signature. If our key length conjecture is correct, then each $\diamondsuit C_j$ should be relatively large, since each such coset should approximately model the frequencies of the underlying language. This all motivates the twist algorithm definition introduced by Barr-Simoson \cite{BaSi2015}.

\begin{defn}
Let $\mathcal{M}$ be a text of length $N$. The \textbf{twist algorithm} finds $m \in \mathbb{N}^{+}$ that maximizes the \textbf{twist index}
$$T(\mathcal{M}, m) = \Big(\frac{100}{m}\sum_{j=1}^{m} \diamondsuit C_j\Big).$$
\end{defn}

While the twist algorithm provided a good first step towards a new key length attack, it does have some significant flaws. The biggest issue is the fact that the twist index is increasing as a function of multiples of the actual key length $k$. This fact is proven for certain cases and verified experimentally for all other cases in \cite{MYPC2023}. As a result, the twist index will always make an incorrect prediction, assuming nontrivial multiples of the key length are part of the domain of $T(\mathcal{M},m)$ for a fixed $\mathcal{M}$. This flaw inspired Park--Kim--Cho--Yum to design the twist$^{+}$ algorithm in \cite{PaKiChYu2020}.

\begin{defn}\label{defnTwistPlus}
Let $\mathcal{M}$ be a text of length $N$. The \textbf{twist$^{+}$ algorithm} finds $m \in \mathbb{N}^{+}$ that maximizes the \textbf{twist$^{+}$ index}
$$T^{+}(\mathcal{M}, m) = T(\mathcal{M},m) - \frac{1}{m-1} \sum_{\mu=1}^{m-1}T(\mathcal{M}, \mu).$$
\end{defn}

Park--Kim--Cho--Yum  provide experimental evidence highlighting the twist$^{+}$ algorithm as far more successful than both the index of coincidence and the twist algorithm for a variety of parameters; see Figure 3 in \cite{PaKiChYu2020}. However, they also highlight the fact that the twist$^{+}$ algorithm does become less effective when short key lengths were used on relatively short texts. Furthermore, it is unclear what domain of $m$-values are considered for maximizing both the twist index and the twist$^{+}$ index in these definitions. In particular, if one increases this domain, then the twist$^{+}$ algorithm decreases in success, as highlighted in Figure 1 and Figure 2 of \cite{MYPC2023}. Similar to the twist algorithm, there is an issue with the twist$^{+}$ algorithm predicting a multiple of the key length, though under more specialized parameters. In hopes of constructing a twist-based algorithm that won't predict multiples of the key length and maintain a high level of accuracy even for large domains of $m$-values, Millichap-Yau--Pate--Carns introduced the twist$^{++}$ algorithm in \cite{MYPC2023}. This algorithm finds the $m$-value that maximizes a local change in twist index. 

\begin{defn}
Let $\mathcal{M}$ be a text of length $N$. The \textbf{twist$^{++}$ algorithm} finds $m \in \mathbb{N}^{+}$ that maximizes the \textbf{twist$^{++}$ index}
$$T^{++}(\mathcal{M}, m) = T(\mathcal{M},m) - \frac{1}{2} \Big(T(\mathcal{M}, m-1) + T(\mathcal{M}, m+1)\Big),$$
where $m \in S \subseteq \{2, \ldots, q\}$ and $N = 12q + r$ for quotient $q$ and remainder $r$.
\end{defn}

Note, this third definition highlights the need to specify a domain of potential key lengths to check. In particular, the value $q$ is set as a maximal $m$-value one should consider since for $m > q$, we have $T(\mathcal{M},m) = 100$. Thus, twist  (and twist$^{+}$ and twist$^{++}$) indices will not provide any useful information for such values. Statistics from Figure 1 and Figure 2 in \cite{MYPC2023} show that the twist$^{++}$ algorithm performs exceptionally well under a variety of parameters, including ones where the twist$^{+}$ algorithm drops in accuracy. While the twist$^{++}$ algorithm is the most accurate twist-based algorithm for most parameters (variety of key lengths and text lengths), there are still some specialized conditions under which the twist$^{++}$ algorithm might predict the largest nontrivial divisor of the actual key length; see Section 3 of \cite{MYPC2023} for a discussion on this and examples. 


\subsection{Other Tests}
\label{subsec:OtherTests}

Here, we briefly discuss a few other statistical and quantitative tools that can assist with key length attacks. These tools were also tested as potential features for our ANNs discussed in Section \ref{sec:NN}.

In \cite{Ma1988}, Matthews introduced two basic tools, $H$ and $\Delta$, that are functions of frequencies of individual letters in a ciphertext and both are highly correlated with the key length $k$. Given a text $\mathcal{M}$, the function $H(\mathcal{M})$ sums the percentage frequencies of the seven most common letters in $\mathcal{M}$, and the function $\Delta(\mathcal{M})$ is the difference between the sum of the percentage frequencies of the seven most common letters in $\mathcal{M}$ and the sum of the percentage  frequencies of the seven least common letters in $\mathcal{M}$.  Matthews performed a regression analysis that showed a linear relationship between  $H$ and $k$  and a linear relationship between $\Delta$ and $k$.  While Matthews highlighted $H$ and $\Delta$ as improvements over the IC, both of these tools were only applied to a specific set of key lengths ($k \in \{3, 5, 9, 13, 17, 21\}$) and accuracy rates were still quite low. 

One other tool that could assist with key length attacks is (information) entropy, which measures the amount of ``information'' in a text and was introduced by Claude Shannon in 1948 \cite{Sh1948}. The first order entropy of a text $\mathcal{M}$ is defined as 
$$H_{1}(\mathcal{M}) = -\sum f_i \log_{2}(f_i),$$ 
where $f_i$ is the relative frequency of the $i^{th}$ letter from the underlying alphabet. Higher order entropies can also be calculated by considering frequencies of digrams, trigrams, etc. With regards to key length attacks, information entropy could serve a similar role to the IC, as both are functions of the individual frequencies of a text and ciphertexts encrypted with varying lengths will have varying frequency distributions.  We refer the reader to Chapter 11 of \cite{Ba2021} for a further introduction to information entropy. 


\section{A Neural Networks Approach to Key Length}
\label{sec:NN}

\par
The discussion in the previous sections illustrates that existing techniques have various strengths and weaknesses depending on a variety of parameters. For example, the twist$^{+}$ algorithm sometimes has trouble distinguishing the correct key length from its multiples, while the twist$^{++}$ algorithm sometimes has trouble distinguishing the correct key length from its largest divisor. Thus, one might want to consider both of these key length tests together along with several other tests, rather than just one of them. A neural network is a natural candidate as a method to combine the existing techniques into one key length finding structure such that the strengths of one technique could potentially compensate for weaknesses in another. In this section, we first give some brief background on Artificial Neural Networks (ANNs) in \ref{subsec:BackgroundNN} and then discuss the specifics of our ANN in \ref{subsec:DataforModel}, \ref{subsec:CreatingModel} and \ref{subsec:model_evaluation}.

\par


\subsection{Background on Neural Networks}
\label{subsec:BackgroundNN}

\par
We first give a brief overview of Feedforward Neural Networks (FFNN) and introduce some of the essential terminology for our work. We direct the reader to \cite{nielsen_2019} and \cite{Goodfellow-et-al-2016} for further background  on ANNs.
\par
A Feedforward Neural Network is a machine learning framework inspired by biological neural networks existing in animal brains. They are both the simplest structure and basis of many machine learning architectures in the class of ``deep learning" algorithms.
\par
\begin{figure}[h]
    \centering
    \begin{tikzpicture}[
  neuron/.style={circle, draw, minimum size=1.5em, inner sep=0pt},
  input/.style={neuron, fill=blue!30},
  hidden/.style={neuron, fill=green!30},
  output/.style={neuron, fill=red!30},
  arrow/.style={-stealth}
]

\foreach \name / \y in {1,...,4}
  \node[input] (I-\name) at (0, 5-\y) {$x_{\name}$};

\foreach \name / \y in {1,...,5}
  \node[hidden] (H-\name) at (2, 5.5-\y) {$h^1_{\name}$};

\foreach \name / \y in {1,...,5}
  \node[hidden] (H2-\name) at (4, 5.5-\y) {$h^2_{\name}$};

\foreach \name / \y in {1,...,3}
  \node[output] (O-\name) at (6, 4.5-\y) {$y_{\name}$};

\foreach \i in {1,...,4}
  \foreach \j in {1,...,5}
    \draw[arrow] (I-\i) -- (H-\j);

\foreach \i in {1,...,5}
  \foreach \j in {1,...,5}
    \draw[arrow] (H-\i) -- (H2-\j);

\foreach \i in {1,...,5}
  \foreach \j in {1,...,3}
    \draw[arrow] (H2-\i) -- (O-\j);

\end{tikzpicture}
    \caption{Example of a Feedforward Neural Network}
    \label{fig:FFNN}
\end{figure}
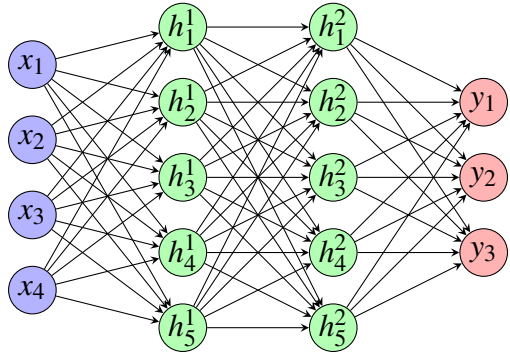

A FFNN is often viewed as a directed, acyclic graph with a number of \textit{layers}. Each vertex of the graph is called a \textit{neuron} and is edge-connected to each neuron in the previous and subsequent layer (other than neurons in the input layer). Figure \ref{fig:FFNN} illustrates a FFNN with input layer consisting of four neurons (this is the left most layer), two hidden layers each with five neurons and an output layer consisting of three neurons.

To each neuron $n$ we associate two numbers: $a_n$, the \textit{activation} of neuron $n$ and a number $b_n$ called a \textit{bias}. To each edge we associate a number called a \textit{weight}, which represents the strength of the connection from one neuron to another. The activation of a neuron $n$, $a_n$ is a function of the activations $a_i$ associated to the neurons in the previous layer connected to neuron $n$, the weights $w_i$ associated to those connections and the bias $b_n$: 

\begin{equation} \label{activation_formula}
    a_n = g(w_1 a_1 + w_2 a_2 + \ldots + w_k a_k + b_n) 
\end{equation}

where $g$ is called an \textit{activation function} and usually has range $[0,1]$. A number of activation functions are used in practice and we refer the reader to \cite[Chapter 6]{Goodfellow-et-al-2016} for further details about these functions.

A FFNN which acts as a classifier can then be seen as a function $f(x; \textbf{W}, \textbf{B})$ with $x$ an input vector of ``features" and parameterized by it's weights \textbf{W} and biases \textbf{B}. For a given  set of weights and biases, when the FFNN is provided an input vector of features, all activations are computed by the formula of Equation \ref{activation_formula}. The output of $f$ is a vector $y$ of activations of the neurons in the output layer, where the entry $y_i$ is the probability that the FFNN believes the input $x$ should be assigned to category $i$. Training the FFNN is the process of determining the optimal weights and biases for doing the job of correctly classifying the inputs for a particular problem, by exposing the network to correctly labelled training examples and evaluating the predictions of the network via a \textit{cost function}. We refer the reader to \cite[Chapter 6]{Goodfellow-et-al-2016}) for further details about training a neural network, including various cost functions. In practice, determining the architecture and weights \textbf{W} (training) of a FFNN is a process of experimenting with different features, activation functions, cost functions, number of layers and number of neurons in each layer.
\par
In this work, we train a FFNN to classify the key length of a  Vigen\`{e}re ciphertext given an input vector of features calculated from the text. We discuss the exact features, activation functions, cost functions, number of hidden layers and neurons used and tested in this work in Section \ref{subsec:CreatingModel}.


\subsection{Data Generation for Model}
\label{subsec:DataforModel}
The data for this project was obtained from the Project Gutenberg website \cite{gutenberg}, an online library of free books. Approximately 5,500 English text files were downloaded, a subset of which were then systematically parsed and cleaned (removing numbers, punctuation and spaces). The cleaned text files were then split into non-overlapping texts of length 200-500. To ensure uniformity of training data, each length $i$, for $200 \le i \le 500$, approximately 1,300 samples were generated. Each sample was then encrypted with random keywords with lengths varying from 3 to 25 characters. The keywords include English words and phrases as well as random strings of letters. Keywords which are English words and phrases were randomly selected from the WordNet database in the Natural Language Toolkit (NLTK) package in Python.
\par
A selection of features based on the key length attacks described in Section \ref{sec:keylength} were then computed for each sample and saved. We discuss the details of these features in Section \ref{subsec:CreatingModel}. Our FFNN was trained on 332,605 samples and then tested on a test set (unseen by the model during the training process) of size 58,695. To do these computations, we used the Tensorflow and Keras libraries in Python.

\subsection{Creating a Model}
\label{subsec:CreatingModel}

We considered a number of features for our ANN based on the techniques discussed in Section \ref{sec:keylength}. The complete list of features we considered in this work are highlighted in Table \ref{fig:initial_features}.

\begin{table}
    \centering
    \begin{enumerate}
    \item Length of ciphertext
    \item Has repeated sequences
    \item Index of coincidence of the ciphertext
    \item Index of coincidence of English (constant, which is 0.066)
    \item The quotient $q$ from $N = 12q + r$ where $N$ is the length of the ciphertext.
    \item The twist indices: $T(\mathcal{M}, m)$ for $1 \le m \le 25$
    \item The twist$^+$ indices: $T^+(\mathcal{M}, m)$ for $2 \le i \le 25$
    \item The twist$^{++}$ indices: $T^{++}(\mathcal{M}, m)$ for $2 \le i \le 25$
    \item The average index of coincidence for cosets $m$ where $3 \le m \le 25$
    \item The $5$ most common distances that occurred between repeated sequences of length $3$ or $4$.
    \item The number of times each of the top 5 most common distances appeared.
    \item Hi-$7$
    \item $\Delta-7$
    \item First order entropy
\end{enumerate}
    \caption{Features considered}
    \label{fig:initial_features}
\end{table}

Let us discuss our rationale for the consideration of these features. Naturally, the experimentally verified effectiveness of the twist$^{+}$ and twist$^{++}$ algorithms in finding the key lengths under various conditions warranted inclusion of  (6), (7) and (8) as features in the network. However, as discussed earlier, when the key length exceeded $N/12$ these tests could not be used. Thus, we included features (5) and (9) to give the network at least some information when this occurred. 
\par
Initial investigations also showed that when repeated trigrams or quadgrams where present, the Babbage-Kasiski test generally yielded useful information. In addition, the Babbage-Kasiski method does not rely on comparing the ciphertext with statistical properties of the English language and hence the inclusion of features (2), (10) and (11). Features (12)-(14) were considered as other possible pieces of information related to the key length as discussed in Section \ref{subsec:OtherTests}.
\par
In order to investigate the relative impact of various subsets of the above features, a standard architecture for the models was initially chosen. We refer to this as the \textit{base NN}. This architecture included an input layer of features coming from Table \ref{fig:initial_features}, two hidden layers of 128 neurons each and an output layer of 23 neurons corresponding with key lengths of $3$ to $25$ characters. Note that, our base NN (and any NN under consideration here) makes a prediction on key length based on whichever of the 23 neurons in the output layer scored the highest probability from this model. We chose ReLu for the activation functions of the hidden layers, and Softmax for the output layer. Our optimizer and loss functions were chosen to be ``Adam" (Adaptive Moment Estimation) and Categorical cross-entropy respectively. These functions were chosen based on fairly standard choices (see \cite[Chapter 6]{Goodfellow-et-al-2016}) for categorization problems. Our training process involved 10 epochs (passes through the training data) and validation test set sizes of 20\% per epoch.
\par
An iterative approach was used to engineer the features of our final neural network. We initiated our investigation of the features by starting with our \textit{base NN} and including all features listed above in Table \ref{fig:initial_features} in Model 1. Based on an analysis of the accuracy of the model (i.e. the proportion of correct predictions) during each epoch of the training process and the accuracy of the model on the unseen test set, we iterated through models by systematically removing each feature (or in some cases, a pair of features) from our original list. At each iteration, if the removal of a feature increased accuracy, further features were removed in the following iteration. When the accuracy rate decreased, we returned those features to the model and removed different features in the next iteration. The results of the feature engineering process are recorded in Table \ref{fig:training_results}. The numbers appearing in the ``Input features" column refers to the numbering of the features in Table \ref{fig:initial_features}.
\begin{table}
    \centering
    \begin{tabularx}{\textwidth}{|X|X|X|X|}
  \hline
  \textbf{Model} & \textbf{Input features} & \textbf{Number of Features} & \textbf{NN Accuracy on unseen test data} \\
  \hline
  Model 1 & All features & 114 &  88.8\% \\
  \hline
Model 2 & All features except 12 and 13 & 112 & 88\%  \\
  \hline
  Model 3 & All features except 14 & 113 & 88.2\% \\
  \hline
  Model 4 & All features except 2 and 14 & 112 & 87.6\%  \\
  \hline
  Model 5 & All features except 14, 3 and 4 &  111 & 87.4\% \\
  \hline
  Model 6 & All features except 14, 5 &  112 & 87.7\% \\
  \hline
  Model 7 & All features except 14, 5, 10 and 11 &  102 & 88.3\% \\
  \hline
  Model 8 & All features except 14, 5, 10, 11 and 9 &  79 & 88.7\% \\
  \hline
  Model 9 & All features except 14, 5, 9, 10, 11 and 6 &  54 & 88.9\% \\
  \hline
  Model 10 & All features except 14, 5, 9, 10, 11, 6 and 7 &  30 & 87.6\% \\
  \hline
  Model 11 & All features except 14, 5, 9, 10, 11, 6 and 8 &  30 & 87.2\% \\
  \hline
\end{tabularx}
    \caption{Model features and accuracy rates}
    \label{fig:training_results}
\end{table}

The feature engineering process is further summarized in Table \ref{fig:feature_importance} below, which displays the effect of leaving out each feature (in the order in which they were removed) on the accuracy rate of the neural network. From Table \ref{fig:training_results} and Table \ref{fig:feature_importance} we see that the best performing model was Model 9, with the following features removed from our original complete feature list: first order entropy, quotient, the Babbage-Kasiski features ((10) and (11) from Table \ref{fig:initial_features}) and the average of the index of coincidences for $m$-cosets.

\begin{table}
    \centering
    \begin{tabular}{|p{0.2\textwidth}|p{0.35\textwidth}|p{0.35\textwidth}|}
    \hline
    \textbf{Model amended} & \textbf{Feature removed} & \textbf{Effect on accuracy rate after removal} \\
    \hline
    Model 1 & Hi-$7$ and $\Delta-7$ & -0.8\% \\
    \hline
    Model 2 & First order entropy & +0.2\% \\
    \hline
    Model 3 & Has repeated sequences & -0.6\% \\
    \hline
    Model 4 & Index of coincidence of the ciphertext and English & -0.2\% \\
    \hline
    Model 5 & The quotient $q$ from $N = 12q + r$ where $N$ is the length of the cipher text. & +0.3\% \\
    \hline
    Model 6 & 10 \& 11 (Babbage-Kasiski features) & +0.6\% \\
    \hline
    Model 7 & The average index of coincidence for cosets $m$ where $3 \le m \le 25$ & +0.4\% \\
    \hline
    Model 8 & Twist indices: $T(\mathcal{M}, m)$ for $1 \le m \le 25$ & +0.2\% \\
    \hline
    Model 9 & Twist$^+$ indices: $T^+(\mathcal{M}, m)$ for $2 \le i \le 25$ & -1.3\% \\
    \hline
    Model 10 & Twist$^{++}$ indices: $T^{++}(\mathcal{M}, m)$ for $2 \le i \le 25$ & -0.4\% \\
    \hline
\end{tabular}
    \caption{Feature importance}
    \label{fig:feature_importance}
\end{table}

Subsequently, we proceeded to further engineer Model 9 by adding back into the model some of the removed features in a different order to test whether different subsets of the removed features could increase model performance. From our experiments, we concluded that adding the average index of coincidences for the $m$-cosets back into the model increased the accuracy rate to 89.2\%. None of the other removed features were able to further improve the performance of Model 9.
\par
In the final step to create our model, we experimented with increasing the number of epochs to 20 during the training process and including an additional layer with 128 neurons in the network. This did not make a significant difference to the accuracy rate. 
\par
Our final model architecture is as follows: Input layer with 77 features, two hidden layers each with 128 neurons and an ouput layer of 23 neurons. 
The activation functions are ReLu for the hidden layers, and Softmax for the output layer. Our optimizer and loss functions are ``Adam" (Adaptive Moment Estimation) and Categorical cross-entropy respectively. The final features included in the neural network are highlighted in Table \ref{fig:features}.
\begin{table}
    \centering
    \begin{enumerate}
    \item Length of ciphertext
    \item Has repeated sequences
    \item Index of coincidence of the ciphertext
    \item Index of coincidence of  English (constant, which is 0.066)
    \item The twist$^+$ indices: $T^+(\mathcal{M}, m)$ for $2 \le i \le 25$
    \item The twist$^{++}$ indices: $T^{++}(\mathcal{M}, m)$ for $2 \le i \le 25$
    \item The average index of coincidence for cosets $m$ where $3 \le m \le 25$
    \item Hi-$7$
    \item $\Delta-7$
\end{enumerate}
    \caption{Final model features}
    \label{fig:features}
\end{table}

\subsection{Evaluation of Model}
\label{subsec:model_evaluation}

\begin{table}
    \begin{center}
\begin{tabular}{|c|c|c|c|c|c|} 
 \hline
 \textbf{Text-Length/Method} & \textbf{IC} & \textbf{Twist Index} & $T^{+}$ & $T^{++}$ & \textbf{Neural Network} \\
 \hline
 200-500 (Overall accuracy) & 7.4\% & 22.4\% & 66.2\% & 63.3\% & 89.2\% \\
 \hline
 200-299 & 6.8\% & 16.1\% & 47.3\% & 42.3\% & 75.1\% \\
 \hline
 300-399 & 7.6\% & 20.3\% & 71.7\% & 65\% & 94.6\% \\
 \hline
 400-500 & 7.7\% & 30.6\% & 79.4\% & 82.2\% & 97.9\% \\
 \hline
\end{tabular}
\end{center}

\caption{Accuracy rates by key length finding method}
    \label{fig:comparison_of_methods_table}
\end{table}


From Table \ref{fig:comparison_of_methods_table}  we see that our neural network is a vast improvement over the index of coincidence (using the formula from \cite[Chapter 3]{Ba2021}) and the twist-based algorithms. In particular, it is able to maintain high accuracy when the ratio of text length to key length is large.
\par
From our original feature list in Table \ref{fig:features}, (14) first order entropy, (5) the quotient $q$ from $N = 12q + r$,(6) the twist indices: $T(\mathcal{M}, m)$ for $1 \le m \le 25$ and (10) and (11) the Babbage-Kasiski features, did not appear to add any value to the neural network, as evidenced by the slight increase in accuracy rate when those respective features were removed. It is perhaps somewhat surprising that the Babbage-Kasiski features were relatively unimportant given that 99\% of the samples contained at least one repeated trigram. One possible explanation could be that to determine the key length from the Babbage-Kasiski data requires an extra layer of intelligent processing to sift through the possible multiples of the key length given in the data and then determine the actual key length from all the possible factors. The twist indices also did not provide useful information in this experiment, which is not surprising given it's overall accuracy rate and the fact that the twist$^{+}$ and twist$^{++}$ indices already implicitly include this information. 
\par
As a rough measure of feature importance, we observed that the removal of the twist$^{+}$ data caused the biggest decrease in accuracy (-1.3\%). We also experimented with removing the twist$^{++}$ data at that step of the engineering process instead, and this yielded a decrease of -1.2\% in accuracy. This suggests that both the twist$^{+}$ and twist$^{++}$ provide strong predictors of key length.
\par
\begin{figure}[H]
    \centering
    \includegraphics[scale=0.46]{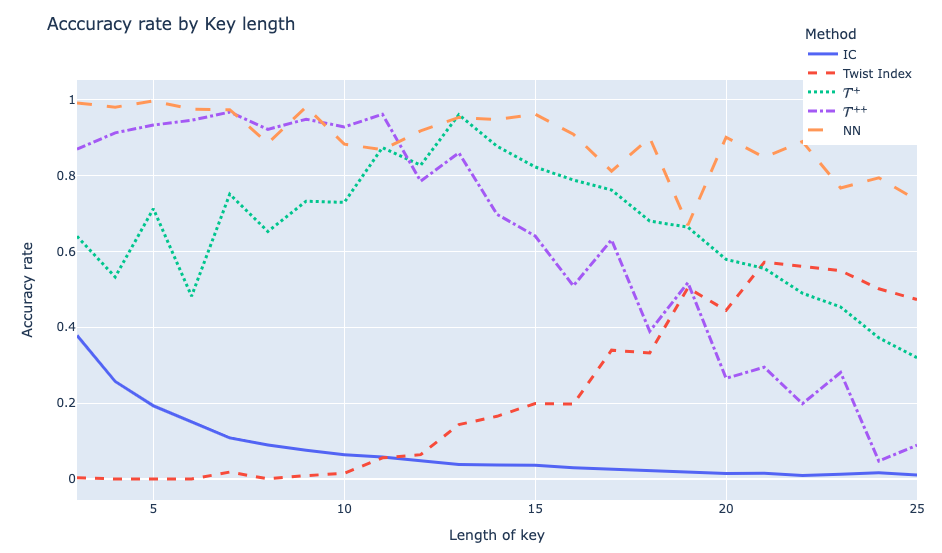}
    \caption{Accuracy rates by key length finding method}
    \label{fig:total_comp_graph}
\end{figure}
\begin{figure}[H]
	\centering
	\includegraphics[scale=0.45]{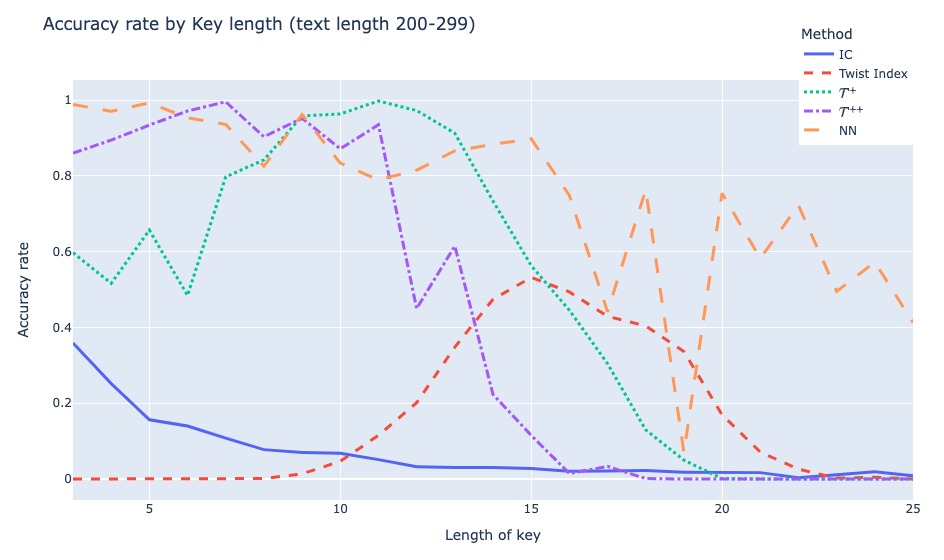}
	\caption{Accuracy rate by key length for $200 \le N \le 299$}
	\label{fig:200_299}
\end{figure}
\begin{figure}[H]
	\centering
	\includegraphics[scale=0.46]{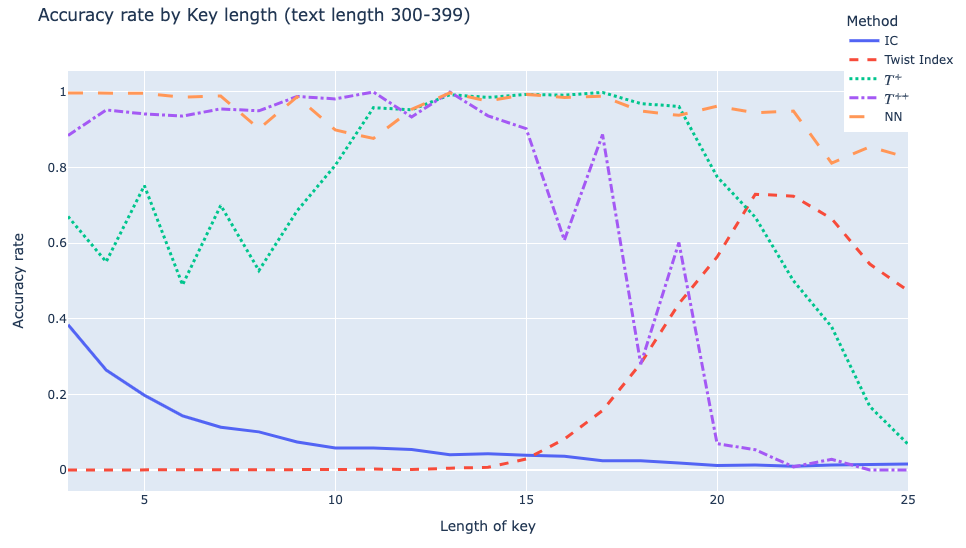}
	\caption{Accuracy rate by key length for $300 \le N \le 399$}
	\label{fig:300_399}
\end{figure}
\begin{figure}[H]
	\centering
	\includegraphics[scale=0.46]{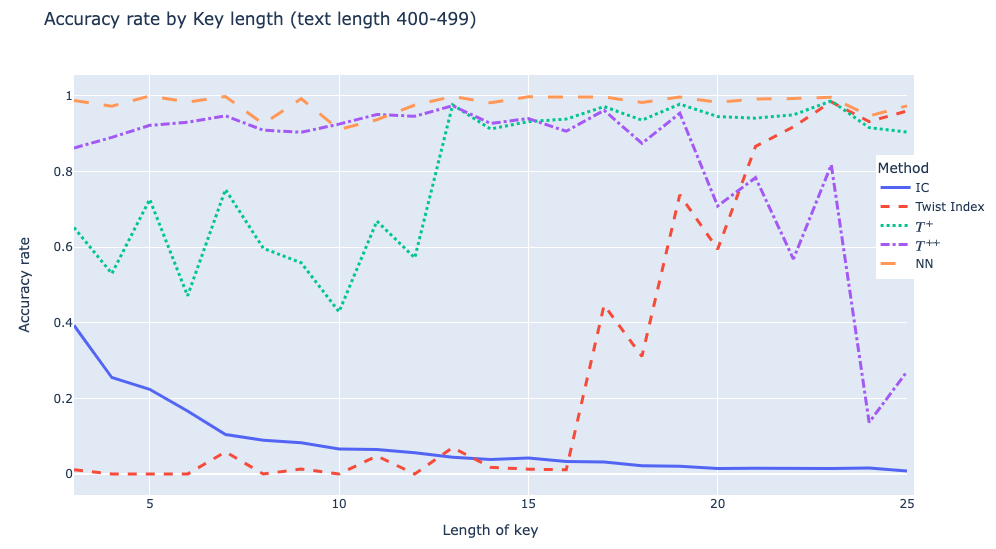}
	\caption{Accuracy rate by key length for $400 \le N \le 499$}
	\label{fig:400_499}
\end{figure}

We now discuss the accuracy of the twist-based algorithms in relation to our neural network, which is highlighted in Figure \ref{fig:total_comp_graph}. In our investigations, the twist$^{+}$ was slightly more accurate than the twist$^{++}$ for longer key lengths, while the twist$^{++}$ was more accurate for shorter key lengths. However, it should be noted that our domain of $m$-values for the twist-based algorithms were restricted to the possible known values of the key length. Since $T(\mathcal{M}, m) \le T(\mathcal{M}, \lambda m)$ for $\lambda \in \mathbb{N}$ (originally stated in \cite{PaKiChYu2020}, with a special case proved in \cite{MYPC2023} along with experimental data verifying this result for the other cases), if a larger range of $m$-values were considered then  we would expect a decline in accuracy in the twist and twist$^{+}$ algorithms, but relatively steady accuracy for the twist$^{++}$. In general, the twist and twist$^{+}$ algorithms frequently predict a multiple of the actual key length when making an incorrect prediction, while the twist$^{++}$ frequently predicts the largest common divisor of the actual key length when making a false prediction. Thus, some of the accuracy ratings in Figure \ref{fig:200_299}, Figure \ref{fig:300_399}, and Figure \ref{fig:400_499} for the twist and twist$^{+}$ algorithms could be overestimates for longer key lengths. In addition, this might mean our neural network's accuracy is slightly overestimated for longer key lengths since the twist$^{+}$ indices are a feature of this network. We refer the reader to \cite{MYPC2023} for a further discussion about the effect of domain size on the accuracy of the twist$^{+}$ and twist$^{++}$ algorithms and most common scenarios for incorrect predictions for these algorithms.

For shorter texts $200 \le N \le 299$ it can be observed from Figure \ref{fig:200_299} that the twist-based algorithms perform very poorly for longer key lengths, but the neural network gives some chance of success. This is due to the fact that each coset in the twist algorithm contains too few letters, resulting in twist indices all being close to 100 (see \cite{MYPC2023} for further discussion).

It is interesting to see that our ANN performs reasonably well under these conditions despite the fact that many of the key features of this ANN individually perform quite poorly under such conditions. 

For longer text lengths $400 \le N \le 500$ we see from Figure \ref{fig:400_499} that the twist$^{+}$ and twist$^{++}$ algorithms are both quite accurate and complement each other in terms of key length, with twist$^{++}$ performing better for shorter key lengths and twist$^{+}$ for longer key lengths. The neural network seems to be able to combine this information and is extraordinarily accurate in these circumstances with an accuracy rate of 97.9\%.

\subsection{Summary}
In this project we investigated how to combine both classical and recent techniques into one key-length finding algorithm via a neural network. We demonstrated that neural networks can be a powerful tool for predicting the key length of a Vigen\`{e}re encrypted text.
\par
We observed that our original model with 114 features was a significant improvement in performance from the twist-based algorithms, and overall, the feature engineering experiment did not make a significant improvement to the success rate of the model, providing some indication that appropriate features were originally chosen.
\par
We were able to feature engineer the network to reduce the number of features in half whilst slightly improving the accuracy of the network, to an overall success rate of 89.2\%. In particular, the neural network is much more accurate than existing methods in predicting key length when the ratio of text length to key length is large. This project also revealed that the recent twist-based algorithms, the  twist$^{+}$ and twist$^{++}$ algorithms, provided some of the strongest indicators of key length.

\section{Biographical Note}

\noindent \textbf{Christian Millichap} is an Associate Professor of Mathematics at Furman University in Greenville, SC. His research interests are in geometric topology and knot theory. He has also enjoyed teaching a variety of classes in cryptology for high school students and undergraduates. 
\\\\
\textbf{Yeeka Yau} is currently a Mathematics Learning Success Advisor at the University of Sydney, Australia. He was recently an Assistant Professor of Mathematics at the University of North Carolina Asheville and a Visiting Assistant Professor at Furman University. His research interests are in Coxeter groups, combinatorial and geometric group theory, historical cryptology and machine learning.

\section{Data Availability Statement}

The data that support the findings of this study are openly available in Zenodo at \\ \href{https://doi.org/10.5281/zenodo.10363051}{https://doi.org/10.5281/zenodo.10363051}.


\section{Acknowledgements}
This work was financially supported by the Furman University Department of Mathematics  via the Summer Mathematics Undergraduate Research Fellowships.

\bibliographystyle{hamsplain}
\bibliography{biblio}

\end{document}